\documentclass[onecolumn,showpacs,preprintnumbers,amssymb]{revtex4}
\usepackage{graphicx}

\def\simlt{\stackrel{<}{{}_\sim}}

\def\beq{\begin{equation}}
\def\eeq{\end{equation}}
\def\bea{\begin{eqnarray}}
\def\eea{\end{eqnarray}}
\def\bit{\begin{itemize}}
\def\eit{\end{itemize}}

\def\nn{\nonumber}

\def\la{\lambda}
\def\La{\Lambda}

\makeatletter
\def\underbracket{%
  \@ifnextchar [ %
    {\@underbracket}%
    {\@underbracket [\@bracketheight]}}
\def\@underbracket[#1]{%
  \@ifnextchar [ %
    {\@under@bracket[#1]}%
    {\@under@bracket[#1][0.4em]}}
\def\@under@bracket[#1][#2]#3{
  \mathop {%
    \vtop {%
      \m@th \ialign {%
        ##\crcr $\hfil \displaystyle {#3}\hfil $%
       \crcr \noalign %
       {\kern 3\p@ \nointerlineskip }%
        \upbracketfill {#1}{#2}
       \crcr \noalign %
       {\kern 3\p@ }%
     }%
   }%
  }%
  \limits%
}
\def\upbracketfill#1#2{%
  $\m@th \setbox \z@ \hbox {$\braceld$}
  \edef\@bracketheight{\the\ht\z@}\bracketend{#1}{#2}
  \leaders \vrule \@height #1 \@depth \z@ \hfill
  \leaders \vrule \@height #1 \@depth \z@ \hfill%
  \bracketend{#1}{#2}$%
}
\def\bracketend#1#2{\vrule height #2 width #1\relax}

\begin{document}
\title{Softly broken conformal symmetry\\[1mm]
and the stability of the electroweak scale}

\author{Piotr H. Chankowski$^1$, Adrian Lewandowski$^1$,
Krzysztof A. Meissner$^{1}$ and Hermann Nicolai$^2$}
\affiliation{$^1$ Faculty of Physics,
University of Warsaw\\
Ho\.za 69, Warsaw, Poland\\
$^2$ Max-Planck-Institut f\"ur Gravitationsphysik
(Albert-Einstein-Institut)\\
M\"uhlenberg 1, D-14476 Potsdam, Germany\\
}

\begin{abstract}
We point out a novel possible mechanism by which the electroweak hierarchy
problem can be avoided in the (effective) quantum field theory. 
Assuming the existence of a UV complete underlying fundamental theory 
and treating the cutoff scale $\La$ of the effective field theory as a
real physical scale we argue that the hierarchy problem would be solved
if the coefficient in front of quadratic divergences vanished for some
choice of $\La$, and if the effective theory mass parameters fixed at
$\La$ by the fundamental theory were hierarchically smaller than $\La$
itself. While this mechanism most probably cannot work in the Standard
Model if the scale $\Lambda$ is to be close to the Planck scale, we
show that it can work in a minimal extension (Conformal Standard Model)
proposed recently for a different implementation of soft conformal
symmetry breaking.
\end{abstract}
\maketitle


The problem of stability of the electroweak scale with respect to the
Planck scale (the so-called hierarchy problem) has for almost 40 years
been one of the main driving forces of theoretical research in high
energy physics. Over the years various mechanisms for solving it 
at the effective quantum field theory level
have been proposed and investigated in detail, of which the most notable 
are technicolor and low energy supersymmetry. With the discovery of
a spin-zero particle at the LHC, and after establishing its basic
characteristics, it has become clear that a solution which departs
little from the simplest mechanism of the electroweak symmetry breaking
realized in the Standard Model (SM) may be preferred. In particular,
extensions of the SM which predict only elementary scalars and no new
higher spin particles other than right-chiral neutrinos seem
distinguished at present. It is therefore of interest that there
exists an alternative way (which does not require new spin $s\geq\frac12$ 
degrees of freedom) by which the problem of stability of the 
electroweak scale could be avoided in the low energy effective theory.
It is based on a novel implementation of `near conformal symmetry' 
in the effective low energy theory.
\vskip0.2cm

As is well known, the classical conformal symmetry of the SM is spoiled
only by the scalar field mass term necessary to induce phenomenologically
viable electroweak symmetry breaking. Moreover, as in any generic quantum 
field theory, conformal symmetry of the SM is broken by quantum effects. 
Yet, the idea that `softly broken conformal symmetry' (SBCS) might be 
relevant for the solution of the hierarchy problem was expressed already 
long ago \cite{BA}. As one possible concrete implementation of this idea 
a minimal extension of the SM, the Conformal Standard Model (CSM), has been  
proposed in \cite{MN1}. Besides the known particles this model only involves 
the right-chiral neutrinos and one extra (complex) scalar field. Originally 
it was assumed that its conformal symmetry is broken only by the anomaly, 
inducing electroweak symmetry breaking via the Coleman-Weinberg mechanism 
\cite{CW}. However, although there do exist perturbatively stable minima 
of the potential of this model giving rise  to a Higgs mass equal to 125 
GeV (as we have checked by carefully investigating the 2-loop effective 
potential of the model), the mixing with the second heavier spin-zero 
particle in all cases turned out too large to be in agreement with 
the LHC data. For this reason, and because of another serious drawback of 
this implementation (related to quadratic divergences, see below) we here 
propose a different way in which SBCS can be at work to solve the hierarchy 
problem, and show how this mechanism can be realized in the model 
\cite{MN1,MNL} with explicit small mass parameters. We also note some 
similarities with the scheme proposed in \cite{SW} in the framework of 
the asymptotic safety program.
\vskip0.2cm

Let us first define our framework. We assume that there exists a complete 
and UV finite fundamental theory (which is likely not a  quantum field theory) 
describing all interactions including (quantum) gravity which,
after integrating out all degrees of freedom above some large scale
$\La$ (presumably close to the Planck scale $M_{\rm Pl}$), fixes the
`bare' action of the effective field theory. In particular, we assume that 
the fundamental theory determines the way the cutoff $\La$ should be 
implemented in the effective theory loop calculations. 
To understand our proposal how the stability of electroweak scale 
at the level of the effective quantum field theory can be secured by
the putative fundamental theory it is crucial to 
keep in mind that, unlike the usual renormalization program
in which $\La$ is eventually taken to infinity, here $\La$ is finite;
for this reason all `bare' parameters of the effective theory fixed at
this scale are also finite. In general the cutoff $\La$ is {\it a priori}
arbitrary: given an UV finite fundamental theory it should always be
possible to integrate out all (gravitational and matter) degrees of
freedom above the scale $\La$ to obtain a finite `bare' effective
theory valid for all energy scales below $\La$.
Even if the fundamental theory does correctly predict (as we
assume) the very small ratio $M^2_{\rm EW}/M^2_{\rm Pl}$ and related
low energy observables, 
that is, even if it completely solves the conceptual aspect of
the hierarchy problem, the effective theory generically is still susceptible
to `technical'  aspect of the problem if it involves scalar fields: if the effective
theory is solved (perturbatively or not)
directly in terms of the bare parameters defined at the scale $\La$,
such small ratios arise as the  result of very precise cancellations of
$\La^2$ contributions against (bare mass)$^2$ parameters of the same order.
\vskip0.2cm

From this perspective the implementation of SBCS as proposed in \cite{MN1}
(as well as in any other model that relies on radiative symmetry breaking
{\it \`a la} Coleman-Weinberg) suffers from the same problem: the absence
of $\La^2$ divergences in the dimensional regularization scheme used there
is, in fact, artificial: in terms of bare parameters, there is a huge
cancellation between the $\La^2$ contributions induced by
real fluctuations of the quantum fields and the (bare mass)$^2$ terms of
the effective action fixed at $\La$ by the fundamental theory which is 
supposed to produce vanishing or very small mass values at the level of 
the effective action.
\vskip0.2cm

Within this general framework one can envisage two different ways in 
which the hierarchy problem at the level of the effective field theory
can be avoided. The first possibility is that the bare parameters 
$m_B^2(\La)$ of the effective theory are hierarchically smaller
than $\La$ {\it and} loop corrections to masses of light particles
proportional to $\La^2$ cancel exactly by some symmetry. This
mechanism is realized in supersymmetric theories \cite{FOOTNOTE}.
In this case the precise value of the cutoff $\La$ does not matter:
the cancellation of the quadratic divergences holds automatically
for {\em any} choice of $\La$. For practical purposes one can then formally
send $\La$ to infinity and adopt any convenient regularization in order
to set up the standard renormalized perturbative expansion.

\vskip0.2cm
The second and novel possibility which we want to point out here is that 
the putative fundamental theory singles out a particular scale $\La$,
the {\em physical} cutoff, at which $m_B^2(\La)\ll \La^2$ {\it and} at which
the complete $\propto\La^2$ corrections to the physical spin-zero boson(s)
(and thus to the ratio $M^2_{\rm EW}/M^2_{\rm Pl}$) vanish.
Naturally one expects $\La$ to be close to the Planck mass $M_{\rm Pl}$.
We will argue below that this can also be regarded as a solution of the 
`technical'  hierarchy problem. Both mechanisms of avoiding the 
hierarchy problem in the effective quantum field theory can thus 
be attributed to SBCS, by small mass terms and by the quantum anomaly.
We stress that neither of these mechanisms solves the `conceptual aspect' 
of the hierarchy problem which probably cannot be solved without knowing
the underlying fundamental theory. However, we point out that since both
the effective theory parameters  at the scale $\Lambda$ as well as the 
prescription how the cutoff $\Lambda$ should be implemented are 
determined by the same fundamental theory (and thus are necessarily 
correlated with each other) it is not inconceivable, that the latter 
theory singles out a scale $\Lambda$ at which our assumptions are satisfied.
\vskip0.2cm

To see how this second possibility manifests itself in a
bottom-up perspective, it is important to realize that
for this the finiteness of the bare parameters must be preserved by 
keeping the cutoff $\La$ finite (in a way dictated by the fundamental 
theory), and for this reason one is not allowed to use continuation in 
space-time dimension to regularize loop integrals in the effective theory 
calculations. Renormalized running parameters can nevertheless be 
introduced by the usual splitting of the mass parameters
$m^2_B(\La)=m^2_R(\La,\mu)+\delta m^2(\La,\mu)$ and couplings
$\lambda_B(\La)=\lambda_R(\La,\mu)+\delta\lambda(\La,\mu)$,
and by fixing the counterterms involving $\delta m^2(\La,\mu)$ and
$\delta\lambda(\La,\mu)$ in the $\La$-${\rm MS}$ subtraction
scheme in which by definition they absorb only contributions proportional to
$\La^2$ and $\ln(\La^2/\mu^2)$ (the counterterms $\delta m^2$) and
$\propto\ln(\La^2/\mu^2)$ (the counterterms $\delta\lambda$).
Computing physical observables within the effective theory one then finds
the following relation between bare and renormalized parameters
\beq\label{laB}
\la_B(\La)
=\la_R+\sum_{L=1}^\infty\sum_{\ell =1}^L a_{L\ell}~\!
\la_R^{L+1}\left(\ln{\La^2\over\mu^2}\right)^\ell.
\eeq
It follows that $\la_B=\la_R$ for $\mu=\La$, and
\bea\label{mB}
m^2_B(\La)&=&m^2_R\, -\,
\hat{f}^{\rm quad} (\mu,\la_R,\La) \, \La^2
+ \, m_R^2\sum_{L=1}^\infty\sum_{\ell =1}^L c_{L\ell}~\!\la_R^{L}
\left(\ln{\La^2 \over\mu^2}\right)^\ell.
\eea
The crucial fact, which is at the heart of our 
proposal \cite{FOOTNOTE0}
is that the coefficient in front of
$\La^2$
\beq
\hat{f}^{\rm quad} (\mu,\la_R,\La)=\,\sum_{L=1}^\infty\sum_{\ell =0}^{L-1} 
b_{L\ell}~\!\la_R^{L}\left(\ln{\La^2\over\mu^2}\right)^\ell \, ,
\eeq
can be written as a function of the {\em bare} coupling(s) only: from 
the analysis of the $\phi^4$ theory \cite{Fujikawa} (which we assume  
to hold generally) it
follows that the logarithmic dependence on the scale $\mu$ of the
$\La^2$ divergence in (\ref{mB}) is spurious because
\bea
\hat{f}^{\rm quad} (\mu,\la_R,\La)=
f^{\rm quad}\big(\la_B(\La)\big).
\eea
In other words, when corrections to the scalar boson mass are computed
in the perturbative expansion in terms of the renormalized parameters,
only non-logarithmic pieces proportional to $\Lambda^2$ in consecutive
orders of the loop expansion correct the form of the function
$f^{\rm quad}$; logarithms multiplying $\La^2$
contribute only to converting the renormalized couplings $\la_R$ into
the bare ones. Thus, an effective quantum field theory
derived from a complete UV finite fundamental theory 
is free from the (``technical'') hierarchy problem if the condition
\beq\label{f=0}
f^{\rm quad}(\la_B)\,=\,0
\eeq
is satisfied! This condition, which from the bottom-up perspective
looks accidental should, according to our assumptions, be viewed as
a manifestation of the intrinsic working of the underlying fundamental
theory.
\vskip0.2cm

As we do not know the scale $\La$ nor the precise way the cutoff
should be implemented, we adopt here a simple smooth cutoff by replacing
$k^\mu\rightarrow k^\mu \exp\left(-\frac{k^2}{2\La^2}\right),$ for each 
momentum in the (renormalizable part of the) action.  
With this prescription the bottom-up procedure to 
check whether a given theory with $n$ physical spin-zero bosons
is free from the hierarchy problem consists in fixing its renormalized
couplings from fits to the low energy data at $M_{\rm EW}$, and then
evolving them with the RG equations as functions of the scale $\mu$ to 
check whether there exists some scale at which the relevant $n$ functions 
$f^{\rm quad}_k$ for $k=1,\dots,n$ (determined to the appropriate loop order) 
vanish simultaneously.
One may then identify this scale with $\La$ and equate $\la_B$ with $\la_R$
at this scale. For consistency, the couplings of the model should
then satisfy the following additional conditions
over the whole range $M_{\rm EW}<\mu< \La$:
\begin{itemize}
\item
there should be neither Landau poles nor instabilities (manifesting
themselves as the unboundedness from below of the effective potential
depending on the running scalar self-couplings);
\item
all couplings $\la_R(\mu)$ should remain small (for the perturbative approach
to be applicable and stability of the effective potential
electroweak minimum).
\end{itemize}

In the SM there is only one possible quadratic divergence associated with
the Higgs boson. Its vanishing was first conjectured in \cite{Veltman}, but
the SM couplings were taken at the electroweak scale, leading to a wrong
prediction for the top quark mass. The RG evolution of the coefficient in front of this divergence was recently investigated in \cite{HKO,Jones}. This analysis indicates that the SBCS requirements 
are not met in the SM: the zero of coefficient function $f^{\rm quad}$ lies  
around $10^{23}$ GeV
(it is hard to accept that the scale at which the effective 
theory should be constructed is so much above the Planck scale)
and furthermore the
scalar self-coupling $\la_R(\mu)$ becomes negative near $\mu=10^{10}$ GeV,
signaling an instability of the electroweak minimum.
Although these statements depend on the loop order considered,
and also (to a considerable extent!) on the precise value of the
top mass, we conclude that in the SM the hierarchy problem
is most likely not  solved by the SBCS mechanism.

We now show that all the necessary conditions can be satisfied by the CSM
of \cite{MN1,MNL}. With explicit mass terms the potential of this model
reads
\bea
V=m_H^2H^\dagger H+m_\phi^2|\phi|^2
+\lambda_1(H^\dagger H)^2+2\lambda_3(H^\dagger H)|\phi|^2
+\lambda_2|\phi|^4,\nonumber
\eea
where $H=(H_1,H_2)$ is the $SU(2)_{\rm EW}$ doublet and $\phi$ is the
extra gauge singlet. At the minimum $\sqrt2\langle H_i\rangle=v_H\delta_{i2}$,
$\sqrt2\langle\phi\rangle=v_\phi$, and the physical spin-zero particles are
the CP-even $h^0$ and $\varphi^0$, which are mixtures
\bea
\left(\matrix{h^0\cr\varphi^0}\right)=
\left(\matrix{c_\beta&s_\beta\cr-s_\beta&c_\beta}\right)
\left(\matrix{\sqrt2~\!{\rm Re}(H_2-\langle H_2\rangle)\cr
\sqrt2~\!{\rm Re}(\phi-\langle\phi\rangle)}\right),\label{eqn:Mixing}
\eea
with masses $M_h$ and $M_\varphi$, and the CP-odd axion
$a^0=\sqrt2\, {\rm Im}\phi$ \cite{LKM}. We assume that $M_h<M_\varphi$.
The existing experimental results suggest that $|\tan\beta|\simlt0.3$, if
$h^0$ is to mimic the SM Higgs boson (see e.g. \cite{BMF}).

Since there are two scalars in this model, {\em two} equations (\ref{f=0})
must be simultaneously satisfied \cite{
FOOTNOTE2}.
At one loop, the two relevant functions $f^{\rm quad}_k$
are straightforward to determine in terms of bare couplings, {\it viz.}
\bea
16\pi^2f^{\rm quad}_1(\la,g,y)=6\la_1+2\la_3+{9\over4}g_w^2
+{3\over4}g_y^2-6y_t^2\nn\\
16\pi^2f^{\rm quad}_2(\la,g,y)=4\la_2+4\la_3-\sum\limits_{i=1}^3y^2_{{}_{N_i}}.
\label{eqn:fconds}\phantom{aaaaaaaa}
\eea
Here $g_w$ and $g_y$ are the $SU(2)_{\rm EW}\times U(1)_Y$ gauge couplings,
$y_t$ is the top quark Yukawa coupling, and $y_{N_i}$ govern the
Majorana-like couplings of the new scalar to the right-chiral neutrinos.
For simplicity (and without much loss in precision) we neglect all other
SM couplings.  In subsequent work we will show that  higher loop
corrections are indeed small with our assumptions.
As there are parameters of the model which are not fixed at present by
the data, we adopt the following procedure to check if the necessary
conditions can be satisfied: we take the known values of the SM couplings
$g_y$, $g_w$, $y_t$ at the electroweak scale and evolve them using
the one-loop RG equations up to
the (reduced) Planck scale $M_{\rm Pl}=2.4\times 10^{18}$ GeV,
which we assume is the scale $\La$ singled out by the fundamental theory.
At $\La=M_{\rm Pl}$ we chose the values of the couplings $\la_1$,
$y_N$ and determine $\la_2$ and $\la_3$ from the vanishing of the
one-loop functions $f_k^{\rm quad}$ (\ref{eqn:fconds}).
The whole set of couplings is then evolved back down to the electroweak
scale. The necessary one-loop $\beta$-functions are given below
(we use the notation $\tilde\beta\equiv16\pi^2\beta$).
For the scalar self-couplings, we have
\bea
\tilde\beta_{\la_1}&=&24\la_1^2+4\la_3^2
-3\la_1\left(3g_w^2+g_y^2-4y_t^2\right)
+{9\over8}g_w^4+{3\over4}g_w^2 g_y^2+{3\over8}g_y^4- 6 y_t^4  \nn\\
\tilde\beta_{\la_2}&=& 20\la_2^2+8\la_3^2
+2\la_2\sum_{i=1}^3y_{{}_{N_i}}^2-\sum_{i=1}^3y_{{}_{N_i}}^4  \nn\\
\tilde\beta_{\la_3} &=&
\frac{1}{2}\la_3\left\{24\la_1+16\la_2+16\la_3
- \left(9g_w^2+3g_y^2\right)
+2\sum_{i=1}^3y_{{}_{N_i}}^2+12y_t^2 \right\}  \nn
\eea
The beta functions of the remaining couplings read:
\bea
\tilde\beta_{g_w} &=& -\frac{19}{6}g_w^3\;, \ \
\tilde\beta_{g_y}=\frac{41}{6}g_y^3, \ \
\tilde\beta_{g_s} =-7 g_s^3,\nn\\
\tilde\beta_{y_t} &=& y_t\left\{ \frac{9}{2}y_t^2
-8g_s^2  -\frac{9}{4}g_w^2
-\frac{17}{12}g_y^2 \right\}, \nn\\
\tilde\beta_{y_{{}_{N_j}}}\!\!\! &=& \frac{1}{2}{y_{{}_{N_j}}}\left\{
2{y^2_{{}_{N_j}}}+\sum_{i=1}^3y_{{}_{N_i}}^2 \right\}.
\eea
At  the electroweak scale the scalar field mass parameters, whose
$\beta$-functions we give here for completeness
\bea
\tilde\beta_{m_H^2}&=&\left\{12\la_1+6y^2_t
-\left({9\over2}g_w^2+{3\over2}g_y^2\right)
\right\}m_H^2
+4\la_{3} m_\phi^2, \nn\\
\tilde\beta_{m_\phi^2}&=&8\la_{3} m_H^2
+\left\{8\la_{2}+\sum^3_{i=1} y^2_{{}_{N_i}}\right\}m_\phi^2\; ,
\eea
are adjusted to give the required values $v_H=246$ GeV and $M_h=125$ GeV.
The mixing angle $\beta$ defined in (\ref{eqn:Mixing}) as well as the $B-L$ breaking scale $v_\phi$ are then expressed as functions of heavy particles masses ($M_\varphi$ and $m_{N_j}\approx y_{N_j} v_\phi/\sqrt{2}$).
\vskip0.2cm

We have performed a numerical scan over  the values (in the range $0\div2$)
of the couplings $\la_1$ and $y_N$ at the scale $\La$,
rejecting all points for which one of the couplings
$\la_1$, $\la_2$ becomes negative (or $\la_3<-\sqrt{\la_1 \la_2}$) 
between the scales $M_{\rm EW}\leq\mu\leq\La$.
A typical plot of the running couplings $\la_i(\mu)$ and $y_N(\mu)$
is shown in Fig.1. Due to the constraints imposed, only solutions with
{\em negative} values of the mixing angle $\beta$ in the range
$0<|\tan\beta| \simlt0.3$ are found. In Fig.2 we show the predicted 
correlation of the masses $m_N$ of the right-chiral neutrinos (here for 
simplicity assumed to be degenerate)
with the mass of the additional scalar $\varphi^0$ and negative values of 
$\tan\beta$ in the allowed range. The extra scalar $\varphi^0$ can decay 
into the usual SM particles (with small widths \cite{KM3}), but also into 
two or three $h^0$'s (the channels $a^0a^0$ and $h^0a^0a^0$ are also open),
or into the lightest right-chiral neutrinos if this 
is kinematically allowed (for instance, with non-degenerate
neutrino masses, not all of which obey $M_\varphi<2m_N$, unlike in Fig.2).
This produces calculable deviations from the `shadow Higgs' behavior described
in \cite{KM3}. These very distinctive features of the CSM would clearly
allow to discriminate it from other models also predicting new heavy scalar 
particles.

We have also checked that the results shown in Figs.1 and 2 are not
very sensitive to the precise choice of the scale $\La$: for example
for the same values of the masses $M_\varphi$ and $m_N$ varying the scale
$\La$ within one order of magnitude changes the value of $\tan\beta$ by
a few percent at most.
\vskip0.2cm

To summarize: we have proposed a novel mechanism by which 
the  effective quantum field theory can avoid the
hierarchy problem. 
We have shown that the 
CSM of \cite{MN1,MNL} can be consistent with this mechanism 
because there does exist a range of values of its parameters
for which {\em all} SBCS requirements can be
satisfied with the scale $\La$ of the order of the Planck scale.
Remarkably, with $\La$ this high, the CSM may provide a complete
scenario within which all problems of particle physics proper can be 
addressed: strong CP-problem is potentially solved \cite{LKM}, neutrinos 
are naturally massive, non-thermally produced axions can constitute 
dark matter, and baryogenesis can probably proceed through leptogenesis 
(whereas the ultimate explanation of the cosmological constant
problem, dark energy,  and of the mechanism driving inflation must 
be relegated to a more fundamental theory of quantum gravity).
Of course, the real test of the model and of the proposed
SBCS scheme would require the detection of the new scalar particle
$\varphi^0$, the heavy neutrinos and the axion. In the further perspective,
with all the parameters of the model fixed from the low energy data it
should become possible to check whether the coefficients in front of
quadratic divergences indeed vanish, and  to fix the scale
$\La$ at which this occurs.

A detailed account of our results will be given elsewhere.

\vspace{0.2cm}
\noindent{\bf Acknowledgments:} AL and KAM thank the AEI for
hospitality and support during this work.

\onecolumngrid
\vspace{0.3cm}
\begin{figure}[ht]
\centering
\includegraphics[scale=0.8]{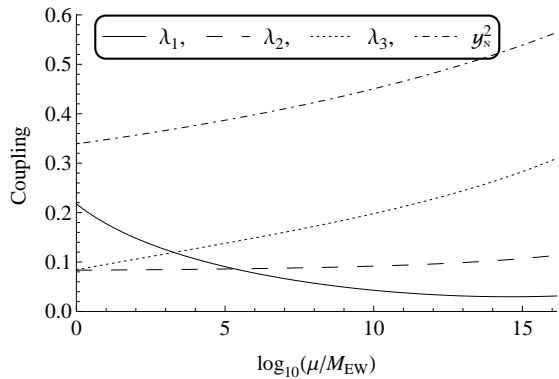}
\label{fig:figure1}
\caption{Running couplings}
\end{figure}
\centering

\begin{figure}[ht]
\centering
\includegraphics[scale=0.8]{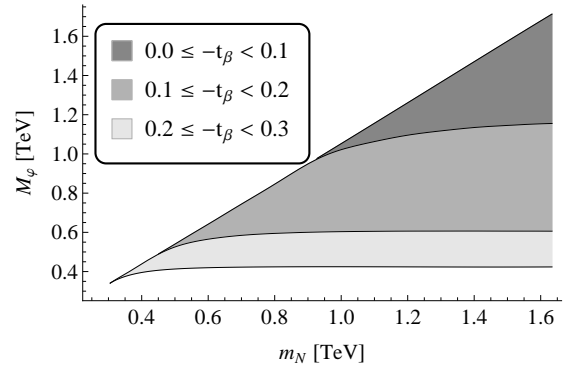}
\label{fig:figure2}
\caption{Predicted correlations of masses $M_\varphi$ with $m_{{}_{N}}$}
\end{figure}

\centering



\begin{thebibliography}{99}
\bibitem{FOOTNOTE} Another possibility, realized in technicolor-type
theories, is that $\La^2$ corrections are absent altogether, and there
are no scalar fields in the effective theory close to $\La$; the Higgs
boson would then be a composite particle.
\bibitem{FOOTNOTE0}The authors of \cite{EJ} did not consider this
possibility because of implicitly taking the cutoff to infinity.
\bibitem{FOOTNOTE2} The mass of $a^0$ which is a (pseudo)Nambu-Goldstone
boson of the spontaneously broken lepton number symmetry is automatically
protected against corrections $\propto\Lambda^2$.
\end{thebibliography}
\end{document}